\documentclass[aps,prd,onecolumn,showpacs,groupedaddress]{revtex4-2}
\usepackage{graphicx}
\usepackage{dcolumn}
\usepackage{bm}
\usepackage{color}
\usepackage{subfigure}
\usepackage{longtable}
\usepackage{supertabular}
\usepackage[T1]{fontenc}
\usepackage{epstopdf}
\usepackage{amsmath}
\usepackage{amssymb}
\usepackage{setspace}
\usepackage{multirow}
\usepackage{booktabs}
\usepackage[section]{placeins}
\usepackage{mathrsfs}
\usepackage{hyperref}
\usepackage{mnsymbol}

\makeatletter

\newcommand{\Rmnum}[1]{\expandafter\@slowromancap\romannumeral #1@} 
\makeatother

\begin{document}
\title{On thermodynamically consistent quasiparticle model at finite chemical potential\footnote{This article is dedicated to the memory of the late Professor Ru-Keng Su.}}

\author{Wei-Liang Qian\textsuperscript{1,2,3}}
\author{Hong-Hao Ma\textsuperscript{4,5}}
\author{Shao-Yu Yin\textsuperscript{6}}
\author{Ping Wang\textsuperscript{7}}

\affiliation{$^{1}$ Escola de Engenharia de Lorena, Universidade de S\~ao Paulo, 12602-810, Lorena, SP, Brazil}
\affiliation{$^{2}$ Faculdade de Engenharia de Guaratinguet\'a, Universidade Estadual Paulista, 12516-410, Guaratinguet\'a, SP, Brazil}
\affiliation{$^{3}$ Center for Gravitation and Cosmology, College of Physical Science and Technology, Yangzhou University, Yangzhou 225009, China}
\affiliation{$^{4}$ Department of Physics, Guangxi Normal University, Guilin 541004, China}
\affiliation{$^{5}$ Guangxi Key Laboratory of Nuclear Physics and Technology, Guangxi Normal University, Guilin 541004, China}
\affiliation{$^{6}$ Institute for Theoretical Physics \& Cosmology, Zhejiang University of Technology, Hangzhou 310032, China}
\affiliation{$^{7}$ Institute of High Energy Physics, CAS, P. O. Box 918(4), Beijing 100049, China}

\begin{abstract}
We explore the quasiparticle model at finite chemical potential related to Ru-Keng Su's distinguished contributions to the topic.
Besides, we discuss recent developments in the model, and in particular, one argues that the effective mass of the quasiparticle might attain a specific form as a function of momentum, in addition to its dependence on temperature and chemical potential.
Unlike the approaches based on the properties of underlying symmetry or renormalization group, the momentum dependence emerges as a special solution to an integro-differential equation resulting from the underlying thermodynamic consistency.
Moreover, this special solution to the problem is shown to be more general than previously explored in the literature.
Instead of fitting to the lattice QCD data at vanishing chemical potential, in this work, we adopt a ``bottom-up'' approach by assuming some analytic ansatzes that are manifestly thermodynamically consistent.
The remaining physical quantities are subsequently derived, and possible implications are also addressed.

\end{abstract}

\date{Dec. 26th, 2022}

\maketitle

\section{Introduction}\label{sec1}

The transition between the quark-gluon plasma (QGP) and hadronic phases constitutes one of the most prominent problems in high-energy nuclear physics.
In the vicinity of such a region, the underlying dynamics are essentially nonperturbative, through which the system undergoes a dramatic change in the number of degrees of freedom.
Moreover, even in the QGP phase, the system's thermodynamic properties deviate significantly from those of a non-interacting ideal gas of quarks and gluons.
For instance, the lattice quantum chromodynamics (QCD) calculations showed~\cite{latt-review-01} that the system's pressure and energy density undershoot the Stefan-Boltzmann limit by about 15-20\% even at temperatures $T \gtrsim 3 T_c$.
Also, the speed of sound extracted from lattice QCD is found to be smaller than that of a massless ideal gas.
In particular, as the system approaches the transition region, it is observed that the speed of sound varies non-monotonically~\cite{qgp-review-03}.
The above properties are crucial for adequately establishing the equation of state (EoS), which plays a central role in providing an appropriate description of the hydrodynamic evolution of the hot and dense system that emerged in the relativistic heavy ion collisions~\cite{hydro-review-10, sph-review-01, hydro-review-09, sph-review-02, sph-eos-02, sph-cfo-01, sph-corr-ev-10, sph-corr-03, sph-corr-04, sph-corr-05, sph-corr-ev-06, sph-corr-ev-08, sph-corr-09}.

In the literature, the density-dependent quark mass was suggested by Fowler, Raha, and Weiner~\cite{quasiparticle-qmdd-01} to address the transition between nuclear and quark matter, and the thermal partonic quasiparticle was initially proposed by Peshier, Kampfer, Pavlenko, and Soff~\cite{quasiparticle-latt-eos-01} to accommodate the numerical experiments from the lattice QCD.
Primarily inspired by its counterparts in other fields of physics, the notion of the quasiparticle is a phenomenological approach aimed at capturing the bulk thermodynamic properties of QGP.
The model can be viewed as an effective and simplified imitation of the essence of many existing theoretical efforts, namely, lattice QCD~\cite{latt-review-01}, dimensional reduction~\cite{qcd-phase-DR-01}, hard thermal loop~\cite{qcd-HTL-01}, Polyakov-loop model~\cite{qcd-phase-PL-01}, as well as other hadronic-degree based approaches~\cite{qcd-phase-Sigma-01}.
It has also been speculated that the success of the notion of quasiparticle degree of freedom will further give rise to novel effective theories from a more fundamental perspective while properly incorporating the nonperturbative aspects of the QCD.

The quasiparticle model interprets the system as a composition of non-interacting quanta which carry the same quantum numbers of quarks and gluons.
The medium-dependent quasiparticle mass implements the strong interactions between the elementary degrees of freedom.
For the description of gluon plasma, the quasiparticle mass was initially assumed to be merely temperature dependent~\cite{quasiparticle-latt-eos-01}.
As the concept flourished in literature, a crucial feature of the model was elaborated in a seminal paper by Gorenstein and Yang~\cite{quasiparticle-gorenstein-01} with respect to its thermodynamic consistency.
The authors solved the issue elegantly via the requirement of an exact cancelation between the additional contributions from the temperature-dependent quasiparticle mass and those from the bag constant.

Subsequently, various relevant aspects of the topic were discussed and further developed by, among others~\cite{quasiparticle-latt-eos-11, quasiparticle-latt-eos-16, quasiparticle-latt-eos-12, quasiparticle-latt-eos-34, quasiparticle-latt-eos-27}, Ru-Keng Su in collaboration with his students and collaborators~\cite{quasiparticle-qmdd-wang-01, quasiparticle-qmdd-zhang-01, quasiparticle-qmdd-zhang-02, quasiparticle-qmdd-zhang-03, quasiparticle-qmdd-zhang-04, phase-04, quasiparticle-qmdd-wu-01, quasiparticle-qmdd-wu-02, quasiparticle-qmdd-wu-04, quasiparticle-qmdd-wu-05, quasiparticle-qmdd-su-03, quasiparticle-qmdd-yin-01, quasiparticle-qmdd-yin-02, quasiparticle-qmdd-yin-03, quasiparticle-qmdd-yin-04, sph-eos-05, sph-eos-06}.
In~\cite{quasiparticle-qmdd-wang-01}, the role of an additional contribution to the thermopotential and its consequential effect on the strange quark matter were explored.
A series of studies regarding the quark mass density- and temperature-dependent (QMDTD) model was performed in~\cite{quasiparticle-qmdd-zhang-01, quasiparticle-qmdd-zhang-01, quasiparticle-qmdd-zhang-02, quasiparticle-qmdd-zhang-03, quasiparticle-qmdd-zhang-04}.
The temperature dependence of the stable radius of a strangelet was discussed in~\cite{quasiparticle-qmdd-zhang-01}.
The temperature dependence of the bag constant $B$ was explored and shown to cure the divergence that occurred at vanishing baryon density in the phase diagram for the bulk strange quark matter of the original QMDTD model~\cite{quasiparticle-qmdd-zhang-02}.
A systematic analysis regarding the stability of strangelet was performed in~\cite{quasiparticle-qmdd-zhang-03} in the framework of the QMDTD model.
It was observed that stable strangelets are more likely to be encountered in the region with a sizeable negative electric charge and significant strangeness.
The analysis was then extended to the dibaryon systems~\cite{quasiparticle-qmdd-zhang-04} regarding different decay channels, and the results were found in good agreement with those obtained by the chiral SU(3) quark model. 
The QMDTD setup was then applied to the context of Friedberg-Lee soliton bag~\cite{friedberg-lee-01, friedberg-lee-02, friedberg-lee-03} nonlinear coupled to the sigma~\cite{quasiparticle-qmdd-wu-01} as well as omega~\cite{quasiparticle-qmdd-wu-02} mesons.
The model was further extended to investigate the properties of deconfinement~\cite{quasiparticle-qmdd-su-03, quasiparticle-qmdd-wu-04} and nuclear matter~\cite{quasiparticle-qmdd-wu-05}.
As an alternative approach to address the thermodynamic consistency, an additional fictitious degree of freedom was introduced~\cite{quasiparticle-qmdd-yin-03, quasiparticle-qmdd-yin-04} to elaborate a generalized version of the first law of thermodynamics.

From the field theory perspective, the mass of a particle can be defined either by the pole of the effective propagator or via the Debye screen mass extracted at small momentum, provided the question of gauge invariance is adequately dealt with.
In particular, the calculations with hard thermal loop approximation show that the gluon screen mass extracted from the above pictures is consistent~\cite{qcd-HTL-01, qcd-HTL-08}.
The derived quasiparticle mass, in turn, is a function of temperature and chemical potential.
As a result, the above dependence calls for a generalization scheme for thermodynamic consistency.
Further developments by Peshier {\it et al.} gives rise to a flow equation~\cite{quasiparticle-latt-eos-11, quasiparticle-latt-eos-16, quasiparticle-latt-eos-12, latt-eos-hydro-16}.
The latter is a partial differential equation, and its boundary condition is chosen at vanishing baryon density, adapted to the lattice QCD data.
It was shown that the thermodynamic properties obtained from such a framework agree well with the lattice calculations at finite baryon chemical potential.

Following~\cite{quasiparticle-gorenstein-01}, one takes the grand partition function of the system ${\cal Z}$ as the starting point, which reads
\begin{eqnarray}
{\cal Z}\left(V, T, \mu\right)=\mathrm{Tr}\left[e^{-\beta \left(H-\mu N\right)}\right]  , \label{gPart}
\end{eqnarray}
and $\beta$ is the reciprocal of the temperature $T$, $\mu$ represents the chemical potential, $V$ is the volume of the system,
$H$ and $N$ are the Hamiltonian and conserved number operators.
In order to derive the remaining thermodynamic quantities (such as pressure, energy density, and conserved number density) in a consistent fashion, the two following identities need to be valid,
\begin{eqnarray}
\frac{\partial {\cal Z}\left(V, T, \mu\right)}{\partial \beta}=-\mathrm{Tr}\left[\left(H-\mu N\right)e^{-\beta \left(H-\mu N\right)}\right]  , \label{therCons01}
\end{eqnarray}
and
\begin{eqnarray}
\frac{\partial {\cal Z}\left(V, T, \mu\right)}{\partial \mu}=\beta\mathrm{Tr}\left[N e^{-\beta \left(H-\mu N\right)}\right]  . \label{therCons02}
\end{eqnarray}

The conditions Eqs.~\eqref{therCons01} and~\eqref{therCons02} are manifestly satisfied when the Hamiltonian is not medium dependent.
As an example, for the quasiparticle model proposed in~\cite{quasiparticle-latt-eos-01},
\begin{eqnarray}
H = \sum_\mathbf{k} \omega(\mathbf{k}, T, \mu) a_\mathbf{k}^\dagger a_\mathbf{k} + V B ,
\end{eqnarray}
where $B$, the {\it bag constant}, is attributed to the vacuum energy, mostly viewed as a constant, and
\begin{eqnarray}
\omega(\mathbf{k}, T, \mu)=\omega(\mathbf{k}, m) =\sqrt{k^2+ m^2} ,\label{defOmega}
\end{eqnarray}
where $k=|\mathbf{k}|$ and $m=m(T)$ is an explicit function of the temperature.
The latter adds an additional contribution to the partial derivative in Eq.~\eqref{therCons01}, associated with $H$.
The recipe by Gorenstein and Yang is derived from the proposal that $B$ should also be medium dependent, namely, $B=B(T)$, whose entire purpose is to identically cancel out the undesirable contribution coming from the temperature-dependent quasiparticle mass.
To be explicit, it is not difficult to show~\cite{quasiparticle-gorenstein-01, quasiparticle-latt-eos-11, sph-eos-06} that the above requirement dictates the relation
\begin{eqnarray}
\frac{dB}{dT} = \left.\frac{\partial p_{\mathrm{id}}(T,\mu,m)}{\partial m}\right|_{T,\mu}\frac{dm}{dT} , \label{go0a}
\end{eqnarray}
where the pressure of ideal gas is an intensive property given by the standard statistical mechanics,
\begin{eqnarray}
p_{\mathrm{id}} = \left.\frac{T}{V}\ln{\cal Z}\left(V, T, \mu\right)\right|_{B=0} , \label{exPress}
\end{eqnarray}
whose specific form is given below in Eq.~\eqref{fpid}.

Since $m=m(T)$ and $B=B(T)$, we have $B=B(m)$.
In other words, Eq.~\eqref{go0a} implies
\begin{eqnarray}
\frac{dB}{dm} = \left.\frac{\partial p_{\mathrm{id}}(T,\mu,m)}{\partial m}\right|_{T,\mu} , \label{go0b}
\end{eqnarray}
where the bag constant $B$ is understood to be a function of the particle mass $m$ only.

Similarly, if the quasiparticle mass is chemical potential dependent, it seems that the above scheme can be readily applied. 
Specifically, one replaces the temperature derivative in Eq.~\eqref{go0a} with the chemical-potential derivative, while Eq.~\eqref{go0b} remains unchanged.
Moreover, if the mass function depends on both the temperature and chemical potential, namely, $m=m(T, \mu)$, Eq.~\eqref{go0b} seemly serves the purpose.
However, though it might not be apparent at first glimpse, one can argue~\cite{sph-eos-06} that Gorenstein and Yang's scheme cannot be applied straightforwardly to such a case.
This can be understood as follows.
To be precise, one needs to solve for $B=B(m)$ for an arbitrarily given form $m=m(T, \mu)$, using Eq.~\eqref{go0b}.
Observing the l.h.s. of Eq.~\eqref{go0b}, one concludes that the dependence of temperature and chemical potential can be entirely ``packed'' into the quasiparticle mass $m$.
On the other hand, since the form $m=m(T, \mu)$ is arbitrary, one can always redefine this function so that the r.h.s. of Eq.~\eqref{go0b} cannot be written as a function of $m$.

We note that the above considerations do not necessarily invalidate Eq.~\eqref{go0b}. 
Instead, it indicates the existence of some additional constraint when finite chemical potential is involved.
In~\cite{quasiparticle-latt-eos-11}, Peshier {\it et al.} derived a flow equation giving a further constraint for the mass function $m=m(T, \mu)$.
In~\cite{sph-eos-06}, some of us derived an integro-differential equation, which is shown to fall back to the flow equation under certain circumstances.
Moreover, it was demonstrated that there are also other possibilities, and in particular, the quasiparticle mass can be a function of the momentum.

In the present study, we proceed further to explore the topic by adopting a ``bottom-up'' approach.
Specifically, instead of numerically adjusting the model parameters to the lattice QCD data, we choose a straightforward but analytical form for the mass function at vanishing chemical potential.
By adopting the analytic function, one can scrutinize the different branches of the mass function in the temperature-chemical potential parameter plane.
The remainder of the present paper is organized as follows.
In the next section, we review the relevant elements regarding the thermodynamic consistency in the quasiparticle model.
The resulting integro-differential equation is presented and discussed.
Sec.~\ref{sec3} focuses on the novel type of solution.
In particular, we explore a mathematically simple form of the mass function at vanishing chemical potential.
It is shown that such a choice will not entirely determine the mass function in the temperature-chemical potential parameter plane.
Different possibilities are then investigated numerically.
The last section is devoted to further discussions and concluding remarks.

\section{The generalized condition for thermodynamical consistency}\label{sec2}

This section discusses the formal constraints for the thermodynamic consistency in the quasiparticle model.
For the present study, the term {\it consistency} implies the following three essential aspects.
First, all the thermodynamic quantities can be derived using the standard formulae once the grand partition function $\cal Z$ is given.
Second, these thermodynamic quantities possess an interpretation in accordance with the ensemble average in statistics.
Last but not least, most thermodynamic identities, for instance, those based on the first law of thermodynamics (c.f. Eq.~\eqref{fLaw}) and extensive properties (c.f. Eq.~\eqref{thId}), remain unchanged.
To our knowledge, the scheme proposed by Gorenstein and Yang is the only one that meets all three above requirements.

As discussed in~\cite{quasiparticle-gorenstein-01}, once Eqs.~\eqref{therCons01} and~\eqref{therCons02} are satisfied, and the energy density and particle number density derived either from the ensemble average or from the partial derivative of the grand partition function possess identical forms.
These lead to the following forms of the energy density
\begin{eqnarray}
\varepsilon=\frac{\langle E\rangle}{V}=-\frac{1}{V}\frac{\partial \ln {\cal Z}}{\partial\beta}= \epsilon_\mathrm{id} + B , \label{energydensity}
\end{eqnarray}
with
\begin{eqnarray}
\epsilon_{\mathrm{id}}=\frac{g}{2\pi^2}\int_0^\infty \frac{k^2dk\omega(k, T, \mu)}{\exp[(\omega(k, T, \mu)-\mu)/T]\mp 1}+\mathrm{c.t.} \ ,
\end{eqnarray}
where $g$ indicates possible degeneracy, ``$\mp$'' corresponds to boson and fermion, and the counter term ``$\mathrm{c.t.}$'' indicates contributions from anti-particles obtained by the substitution $\mu\rightarrow -\mu$ in the foregoing term.
We have also considered the isotropic case $m(\mathbf{k}, T, \mu)=m(k, T, \mu)$.

To derive the above equation, we have assumed the validity of Eq.~\eqref{therCons01}, namely, the contribution from the temperature dependence of quasiparticle mass has precisely been canceled out with the temperature dependence of $B$.
By writing it out explicitly, one finds
\begin{eqnarray}
\frac{\partial B}{\partial T}= -\frac{g}{2\pi^2}\int_0^\infty \frac{k^2dk}{\omega(k ,T,\mu)}\frac{1}{\exp[(\omega(k,T,\mu)-\mu)/T]\mp 1}m\frac{\partial m}{\partial T} + \mathrm{c.t.} \label{go2}
\end{eqnarray}

In statistical mechanics, the pressure is interpreted as a ``general force'', which reads
\begin{eqnarray}
p=\frac{1}{\beta}\frac{\partial \ln {\cal Z}}{\partial V}=\frac{1}{\beta}\frac{\ln {\cal Z}}{V}=p_{\mathrm{id}} - B \label{pressure},
\end{eqnarray}
where
\begin{eqnarray}
p_{\mathrm{id}}&=&\frac{\mp g}{2\pi^2}\int_0^\infty k^2dk\ln \left\{ 1\mp\exp\left[\left(\mu-\omega(k,T,\mu)\right)/T\right]\right\}+\mathrm{c.t.}  \nonumber\\
&=&\frac{g}{12\pi^2}\int_0^\infty \frac{k^3dk}{\exp[(\omega(k,T,\mu)-\mu)/T]\mp 1}\left.\frac{\partial \omega(k,T,\mu)}{\partial k}\right|_{T,\mu}+\mathrm{c.t.} \label{fpid}
\end{eqnarray}

Also, as an ensemble average, the number density is found to be
\begin{eqnarray}
n=\frac{\langle N \rangle}{V}=-\frac{1}{V}\frac{\partial \ln {\cal Z}}{\partial\alpha}=n_{\mathrm{id}} ,\label{fnb}
\end{eqnarray}
where
\begin{eqnarray}
n_{\mathrm{id}}=\frac{g}{2\pi^2}\int_0^\infty \frac{k^2dk}{\exp[(\omega(k,T,\mu)-\mu)/T]\mp 1}-\mathrm{c.t.} 
\end{eqnarray}
Again, we have assumed the condition Eq.~\eqref{therCons02}, which states that the contribution from the chemical-potential dependence of quasiparticle mass in the ideal gas term and that from the bag constant $B$ cancel out each other.
The above condition can be specified to give
\begin{eqnarray}
\frac{\partial B}{\partial\mu}= -\frac{g}{2\pi^2}\int_0^\infty \frac{k^2dk}{\omega(k,T,\mu)}\frac{1}{\exp[(\omega(k,T,\mu)-\mu)/T]\mp 1}m\frac{\partial m}{\partial \mu} + \mathrm{c.t.} \label{go3}
\end{eqnarray}

The well-known thermodynamic identity
\begin{eqnarray}
\epsilon = T\frac{\partial p}{\partial T} - p +\mu n ,\label{thId}
\end{eqnarray}
essentially comes from the first law of thermodynamics and its extensive properties.
As a matter of fact, following the procedure elaborated in the standard textbook~\cite{book-statistical-mechanics-pathria}, it is not difficult to verify that the total derivative of $q=\ln {\cal Z}$ gives
\begin{eqnarray}
dq=-\langle N \rangle d\alpha - \langle E \rangle d\beta - \beta p dV  .
\end{eqnarray}
By comparing the above expression with the first law of thermodynamics, namely,
\begin{eqnarray}\label{fLaw}
d\langle E\rangle=T dS - p dV + \mu d\langle N\rangle  ,
\end{eqnarray}
we have the mapping
\begin{eqnarray}
&&\beta = \frac{1}{k_{B}T} ,  \nonumber \\
&&\alpha = -\frac{\mu}{k_{B}T} ,  \nonumber\\
&&q+\alpha N +\beta E = \frac{S}{k_{B}} .\label{eentropy}
\end{eqnarray}
The validity of Eq.~\eqref{thId} is readily verified.

Now, we proceed to discuss the implications of the conditions Eqs.~\eqref{go2} and~\eqref{go3}.
By taking partial derivative of Eq.~\eqref{go2} in $\mu$ and compare with the partial derivative of Eqs.~\eqref{go3} in $T$, one arrives at the following integro-differential equation~\cite{sph-eos-06}
\begin{equation}
\begin{aligned}
\llangle m\frac{\partial m}{\partial T} \rrangle_- = \llangle m\frac{\partial m}{\partial \mu}\rrangle_+ , \label{gozero}
\end{aligned}
\end{equation}
where
\begin{equation}
\begin{aligned}
\llangle O \rrangle_- &\equiv \int_0^\infty k^2dk\left\{\frac{\exp[(\omega-\mu)/T]}{(\exp[(\omega-\mu)/T]\mp 1)^2 \omega T}-\mathrm{c.t.}\right\}O(k) , \\
\llangle O \rrangle_+ &\equiv \int_0^\infty k^2dk\left\{\frac{\exp[(\omega-\mu)/T](\omega-\mu)}{(\exp[(\omega-\mu)/T]\mp 1)^2 \omega T^2}+\mathrm{c.t.}\right\}O(k) . \label{gozero2}
\end{aligned}
\end{equation}

The solution of Eq.(\ref{gozero}), $m=m(k,T,\mu)$, is in general a function also of the momentum $k$. 
In turn, the bag constant $B$ is obtained by integrating Eqs.~\eqref{go2} and~\eqref{go3} on the parameter plane.
It can be viewed as a functional of $m(k, T,\mu)$ besides being a function of $T$ and $\mu$.
It is noted that the above discussions can be straightforwardly generalized to the case where the system is not isotropic, where $m = m(\mathbf{k}, T,\mu)$.
As pointed out in~\cite{sph-eos-06}, if one simplifies and considers the momentum-independent case, namely, $m(\mathbf{k}, T,\mu) = m(T,\mu)$, one readily falls back to the flow equation derived in Ref.~\cite{quasiparticle-latt-eos-11}.
In this case, $B$ also simplifies to a function of $T$ and $\mu$.
We are, however, more interested in exploring the momentum-dependent case, which will be elaborated further in the following section.

\section{Bottom up toy-model approaches}\label{sec3}

An apparent momentum-dependent solution to Eq.~\eqref{gozero} can be obtained by ``factoring out'' the momentum integration $\int k^2dk$ and assuming the integrand vanishes. 
In other words,
\begin{equation}
\begin{aligned}
\left\{\frac{\exp[(\omega-\mu)/T]T}{(\exp[(\omega-\mu)/T]\mp 1)^2}-\mathrm{c.t.}\right\}\frac{\partial m}{\partial T} =\left\{\frac{\exp[(\omega-\mu)/T](\omega-\mu)}{(\exp[(\omega-\mu)/T]\mp 1)^2}+\mathrm{c.t.}\right\}\frac{\partial m}{\partial \mu} . \label{godown}
\end{aligned}
\end{equation}
The above equation can be solved by using the method of characteristics~\cite{book-methods-mathematical-physics-01}.
To be specific, for a given $k$, the solution is the surface tangent to the vector field 
\begin{equation}
\left(a(T, \mu, m), b(T, \mu, m), 0\right) , \nonumber
\end{equation}
where
\begin{equation}\label{abCo}
\begin{aligned}
a(T, \mu, m) &= \frac{\exp[(\omega-\mu)/T]T}{(\exp[(\omega-\mu)/T]\mp 1)^2}-\mathrm{c.t.} , \\
b(T, \mu, m) &= -\frac{\exp[(\omega-\mu)/T](\omega-\mu)}{(\exp[(\omega-\mu)/T]\mp 1)^2}-\mathrm{c.t.}. 
\end{aligned}
\end{equation}
Its formal solution is the characteristic curves obtained by the integration
\begin{equation}
\begin{aligned}
\frac{dT}{d\lambda} &=a(T, \mu, m) , \\
\frac{d\mu}{d\lambda}&=b(T, \mu, m) ,
\end{aligned}
\end{equation}
where $\lambda$ is an intermediate variable, for given $k$, $m$, and thus $\omega$.

An interesting scenario that gives rise to an analytic solution occurs when one ignores anti-particles' contributions.
By taking $\omega, T, \mu$ as the three independent variable, the method of characteristics gives~\cite{sph-eos-06} the following formal solution
\begin{eqnarray}
m=f\left(\frac{T\omega}{\omega-\mu}, k\right) , \label{go7b}
\end{eqnarray}
where an arbitrary function $f$ furnishes the boundary condition at vanishing chemical potential, namely, $f(T)\equiv f(T,0)= m(T,\mu=0, k=0)$, where we assume that the mass is independent of the momentum at $\mu=0$.
We note that the resultant mass function is a function of $k, T$, and $m$, and therefore the solution of the form Eq.~\eqref{go7b} serves as a simple but non-trivial example.

In~\cite{sph-eos-06}, the freedom in $f$ was employed to perform a numerical fit to the lattice QCD results for $N_f=2+1$ flavor QCD system~\cite{latt-eos-12,latt-eos-14,latt-eos-18,latt-eos-15,latt-eos-19} at vanishing chemical potential.
Then the relevant physical quantities, such as the trace anomaly, sound velocity, and particle number susceptibility, were evaluated and compared to the lattice data.
Instead, we adopt a ``bottom-up'' approach for the present study.
Specifically, we consider two cases where one assumes a simple ansatz for $f$ posteriorly adapted to the lattice results and proceeds analytically to a large extent. 

\begin{figure}[htb]
\centering
\includegraphics[width=0.5\textwidth]{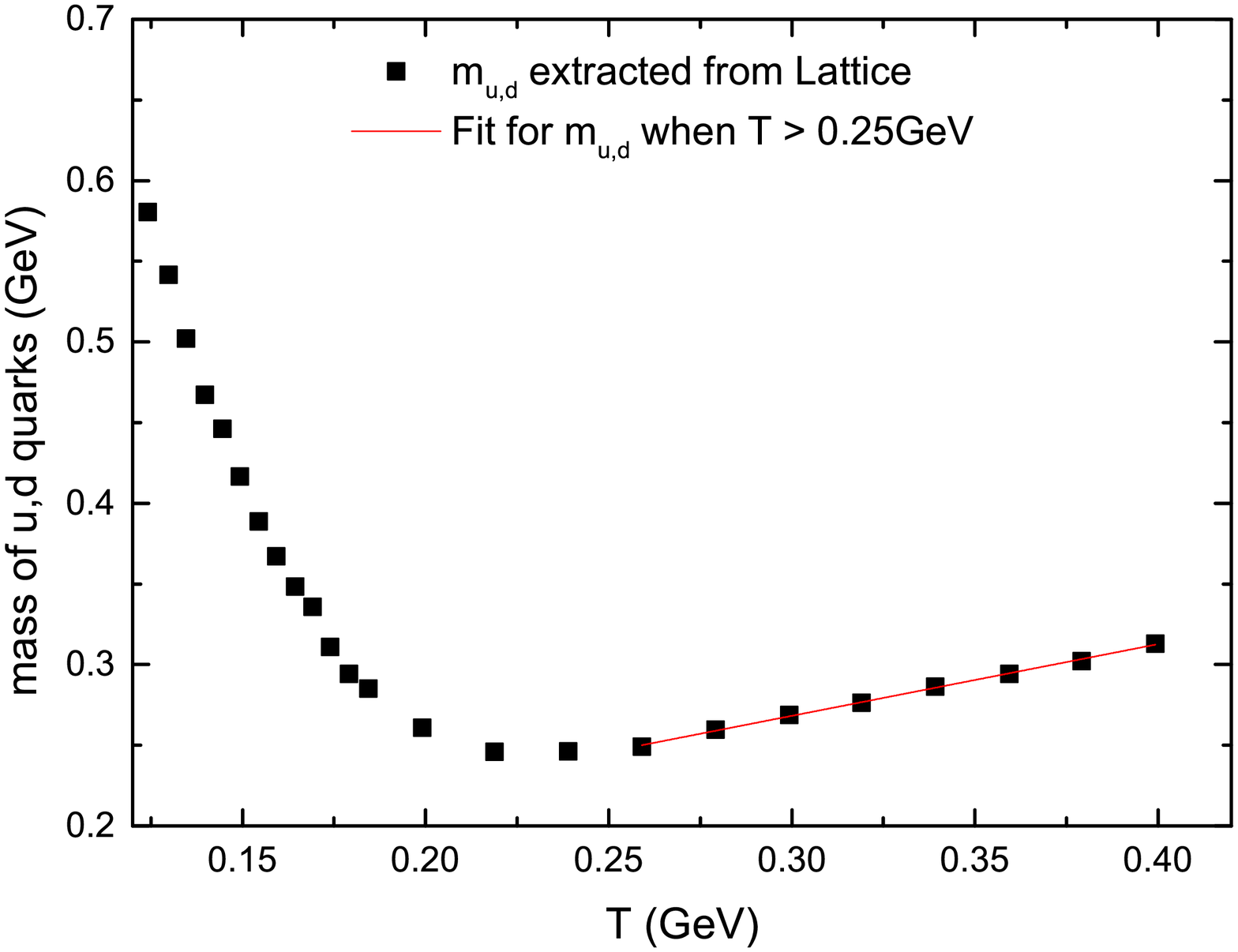}
\caption{The mass of $u$ and $d$ quarks at vanishing chemical potential derived from the lattice data~\cite{latt-eos-12}.
It can be readily extracted for vanishing chemical potential using Gorenstein and Yang's scheme~\cite{quasiparticle-gorenstein-01}.
The curve is then fit to analytic form Eq.~\eqref{ansaL01} discussed in the text.}
\label{massud}
\end{figure}

{\it Case 1:} Our first choice is a simple linear fit.
Based on the lattice data~\cite{latt-eos-12} shown in Fig.~\ref{massud}, there are two regions where the mass of the quasiparticle is primarily a linear function in temperature. 
In other words,
\begin{eqnarray}
\left.f\right|_{\mu=0}=c_1 T + c_2 , \label{ansaL01}
\end{eqnarray}
which gives
\begin{eqnarray}
f\left(\frac{T\omega}{\omega-\mu}, k\right)=f\left(\frac{T\omega}{\omega-\mu}\right)= \frac{c_1 T\omega}{\omega-\mu} + c_2 . \label{ansaL1}
\end{eqnarray}

Despite its simple form, Eq.~\eqref{ansaL01} might be plagued by the pole on its denominator.
To avoid the pole at $\omega=\mu$ for an arbitrary momentum $k$ indicates the condition
\begin{eqnarray}
\omega > \mu , \label{ansaCon1}
\end{eqnarray}
that is, $\omega \ge m > \mu$, by considering the definition Eq.~\eqref{defOmega}.
Otherwise, if one requires $\omega<\mu$, it is always possible to find a momentum $k$ large enough to violate the condition.

Substituting Eq.~\eqref{ansaL1} into Eq.~\eqref{go7b} gives
\begin{eqnarray}
\omega (m-c_1 T- c_2) = \mu (m -c_2) ,\label{Eqcase1}
\end{eqnarray}
for which Eq.~\eqref{ansaCon1} dictates
\begin{eqnarray}
c_1 > 0 , \label{ansaCon2}
\end{eqnarray}
while given $T>0$ and $\mu>0$.

By substituting Eq.~\eqref{defOmega} into Eq.~\eqref{Eqcase1} and squaring both sides, one finds a fourth-degree polynomial equation for $m$.
This equation possesses four roots, where complex roots always appear in pairs.
The physically relevant solution must sit on the positive real axis.

\begin{figure}[htb]
  \centering
  \subfigure[]{
    \includegraphics[scale=0.28]{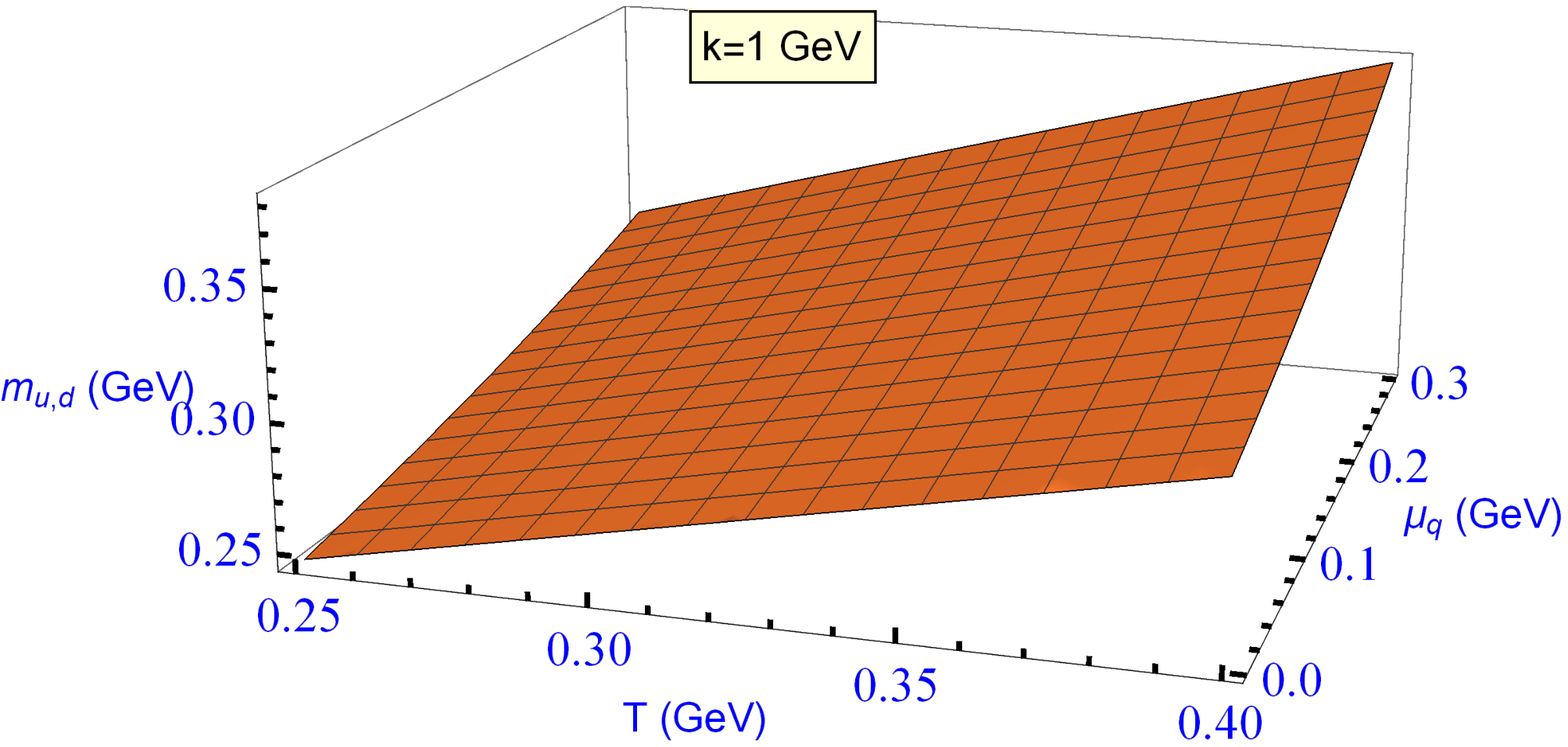}}
  \hspace{0.00in}
  \subfigure[]{
    \includegraphics[scale=0.28]{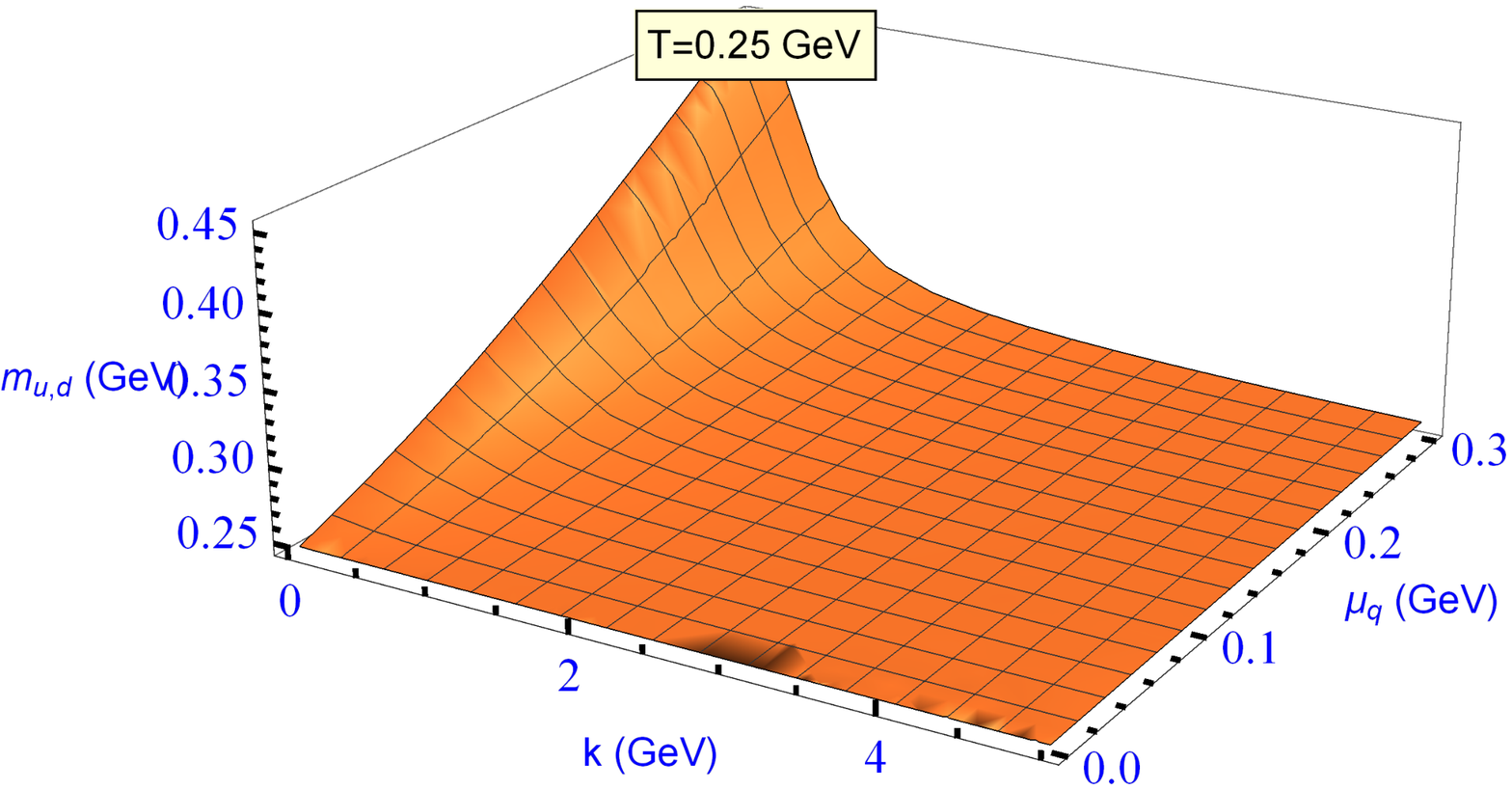}}
  \hspace{0.00in}
  \subfigure[]{
    \includegraphics[scale=0.28]{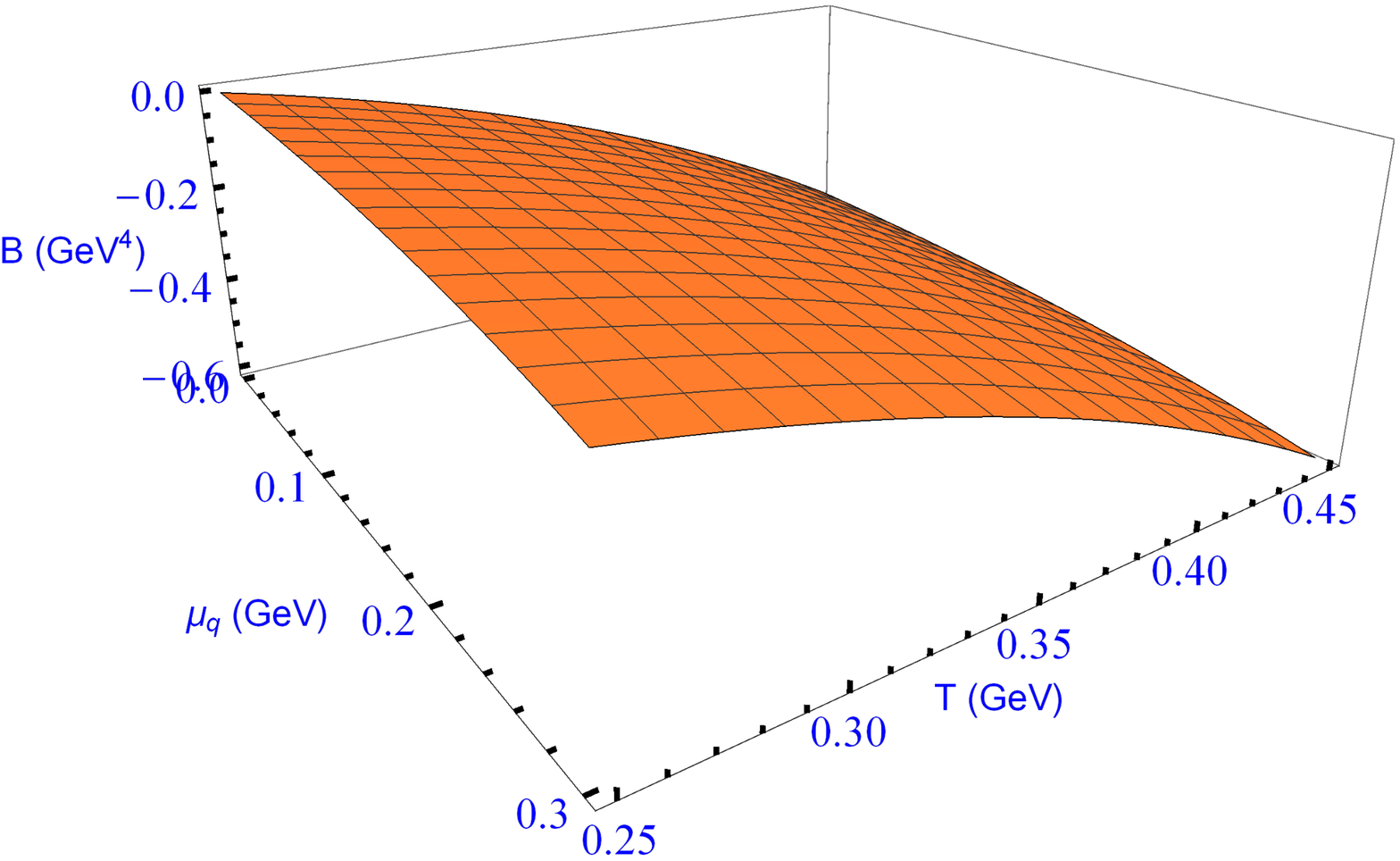}}
\caption{The derived quasiparticle mass in the parameter space according to the form given by Eq.~\eqref{ansaL1} and the fit shown in Fig.~\ref{massud}.
(a) The quasiparticle mass $m$ as a function of $T$ and $\mu$ at $k = 1$ GeV.
(b) The quasiparticle mass $m$ as a function of $k$ and $\mu$ at $T = 0.25$ GeV.
(c) The bag constant $B$ as a function of $T$ and $\mu$.} 
\label{massfinal}
\end{figure}

From this point on, we proceed numerically.
One extracts the values $c_1$ and $c_2$ from the region shown in Fig.~\ref{massud} satisfying $c_1 > 0$, and finds $c_1= 0.44$ and $c_2=0.13$.
The fourth-degree polynomial contains a pair of complex roots, which are subsequently discarded.
One of the remaining two real roots is extraneous, owing to the fact that we have squared both sides of Eq.~\eqref{Eqcase1}.
The resultant mass function $m(k, T, \mu)$ and the bag constant $B(T, \mu)$ are shown in Fig.~\ref{massfinal}.
The left and middle plots give the mass as a function of $T$ and $\mu$ at given $k=1$ GeV and that of $k$ and $\mu$ at given $T=0.25$ GeV.
The resulting bag constant is obtained by numerical integration of Eqs.~\eqref{go2} and~\eqref{go3}.
The dependence of the bag constant on the temperature $T$ or chemical potential $\mu$ is presented in the right plot of Fig.~\ref{massfinal}.
The mass function and the bag constant are found to be moderate in $T$ and $\mu$.
As $k\to 0$, according to the middle plot of Fig.~\ref{massfinal}, the quasiparticle mass increases significantly.
It is noted that the obtained bag constant $B$ is manifestly path independent.
For instance, one evaluates $B(T,\mu)$ by using two following integration paths on the $T-\mu$ plane.
The integration for $B$ is carried out from $(T_0=0.25$ GeV $,\mu_{0}=0)$ to $(T_1=0.45,\mu_1=0.3)$,
where path 1 is defined by $(T_0,\mu_0)\rightarrow (T_1,\mu_0)\rightarrow (T_1,\mu_1)$,
while path 2 is through $(T_0,\mu_0)\rightarrow (T_0,\mu_1)\rightarrow (T_1,\mu_1)$.
One finds 
\begin{eqnarray}
\left.\left[B(T_1,\mu_1)-B(T_0,\mu_0)\right]\right|_{\rm{path~1}}= -0.606049 = \left.\left[B(T_1,\mu_1)-B(T_0,\mu_0)\right]\right|_{\rm{path~2}} .\nonumber
\end{eqnarray}

We also note that based on the above discussions, the fit to the region $c_1 <0$, where the mass of the quasiparticle decreases with increasing temperature in Fig.~\ref{massud}, is doomed to fail.
A numerical attempt reveals path-dependent values, which signals that those obtained by straightforward integration do not yield mathematically well-defined results.
This is due to the undesirable pole at $\omega=\mu$ in the denominator of the first term on the r.h.s. of Eq.~\eqref{ansaL1}.
In order to handle the region where $c_1 <0$, we proceed to consider the second case.

{\it Case 2:} The second choice involves a linear function in the reciprocal of the argument of Eq.~\eqref{go7b}.
To be specific, we consider the ansatz,
\begin{eqnarray}
\left.f\right|_{\mu=0}=\frac{1}{c_3 + c_4 T}, \label{ansaL02}
\end{eqnarray}
which gives
\begin{eqnarray}
f\left(\frac{T\omega}{\omega-\mu}, k\right)= \frac{\omega-\mu}{c_3(\omega-\mu)+c_4 T \omega } . \label{ansaL2}
\end{eqnarray}

In this case, to avoid the pole in the denominator, one considers the following constraint
\begin{eqnarray}
\omega(c_4 T + c_3) > \mu c_3 .\label{ansaCon3}
\end{eqnarray}
Substituting Eq.~\eqref{ansaL2} into Eq.~\eqref{go7b} gives
\begin{eqnarray}
\omega (c_4 T m+ m c_3 - 1) = \mu (m c_3 - 1) .\label{Eqcase2}
\end{eqnarray}
Given $T>0$ and $m>0$, Eq.~\eqref{ansaCon3} implies 
\begin{eqnarray}
\omega>\mu, \ \ c_4>0, 
\end{eqnarray}
and moreover, Eq.~\eqref{Eqcase2} further indicates
\begin{eqnarray}
c_3 < \frac1m, \ \ c_4Tm+m c_3 <1 .
\end{eqnarray}
Otherwise, if $c_3 \ge 1/m$, Eq.~\eqref{Eqcase2} can no longer hold.

To proceed, one substitutes Eq.~\eqref{defOmega} into Eq.~\eqref{Eqcase2} and squares both sides, and one again finds a fourth-degree polynomial equation for $m$.
Similarly, this equation possesses four roots, where complex roots appear in pairs.

\begin{figure}[htb]
\centering
\includegraphics[width=0.5\textwidth]{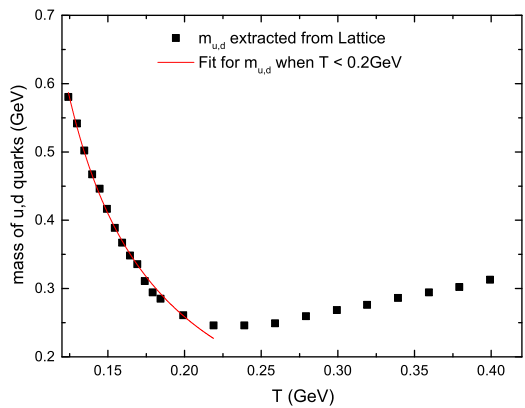}
\caption{The mass of $u$ and $d$ quarks at vanishing chemical potential derived from the lattice data~\cite{latt-eos-12}, which is fit to analytic form Eq.~\eqref{ansaL02} discussed in the text.}
\label{massud2}
\end{figure}

We proceed numerically from this point on.
The values $c_3$ and $c_4$ are extracted from a fit to the lattice QCD data shown in Fig.~\ref{massud2}.
One finds $c_3 = -1.84$ and $c_4 = 28.55$, which affirms the second choice above.
By discussing a pair of complex roots and an extraneous root, the physically relevant solution is eventually singled out from the two sit on the positive real axis.

\begin{figure}[htb]
  \centering
  \subfigure[]{
    \includegraphics[scale=0.28]{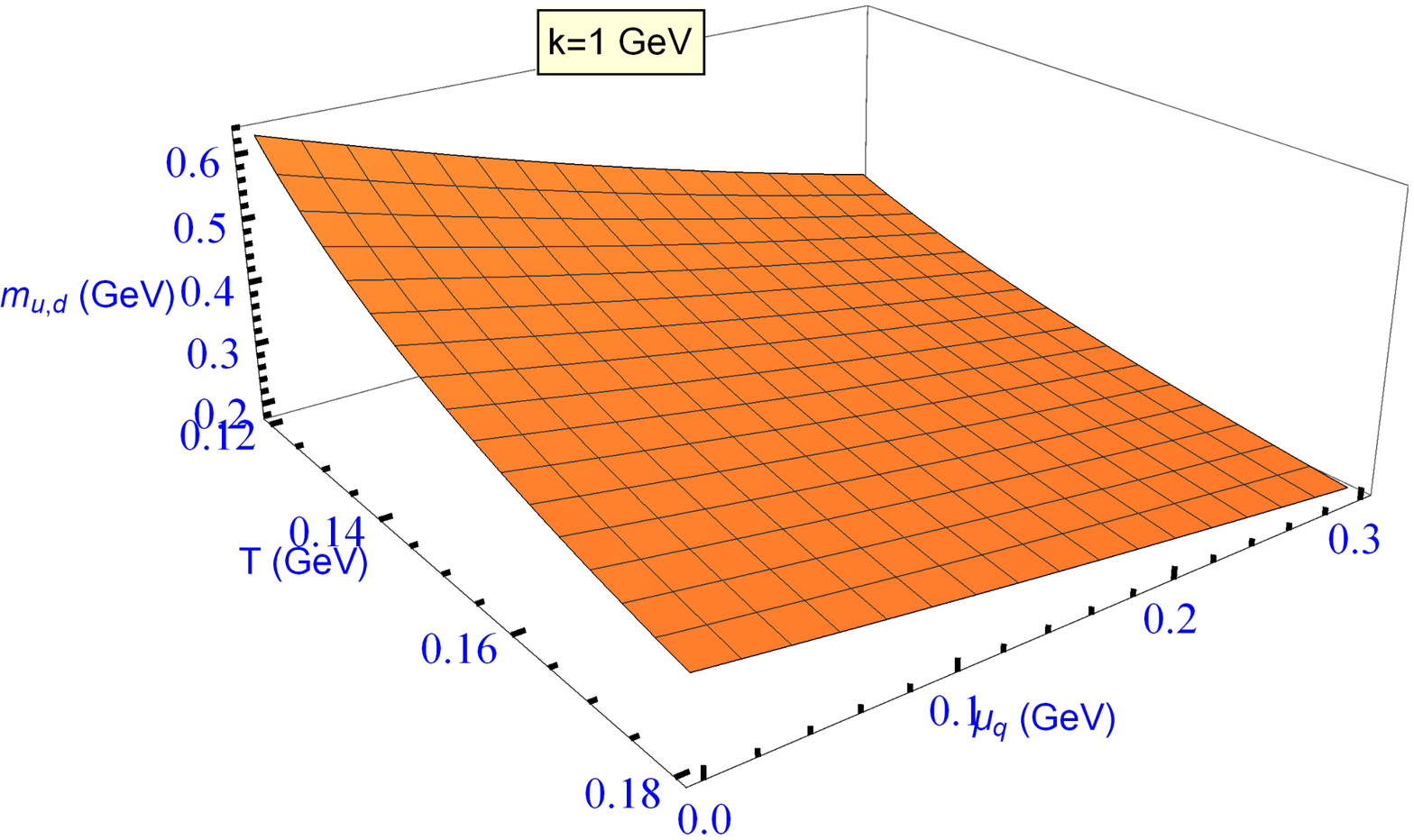}}
  \hspace{0.01in}
  \subfigure[]{
    \includegraphics[scale=0.28]{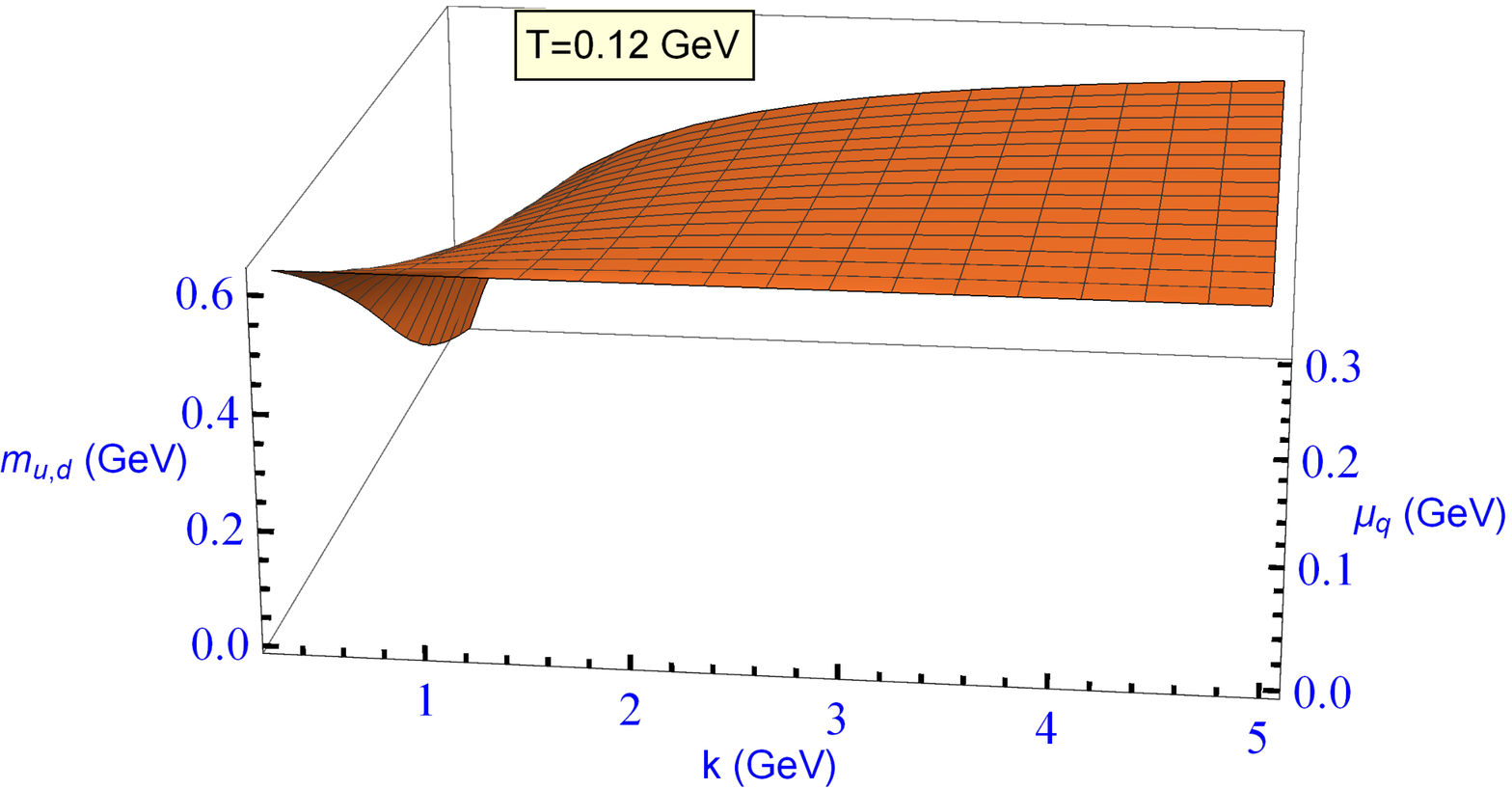}}
  \hspace{0.01in}
  \subfigure[]{
    \includegraphics[scale=0.28]{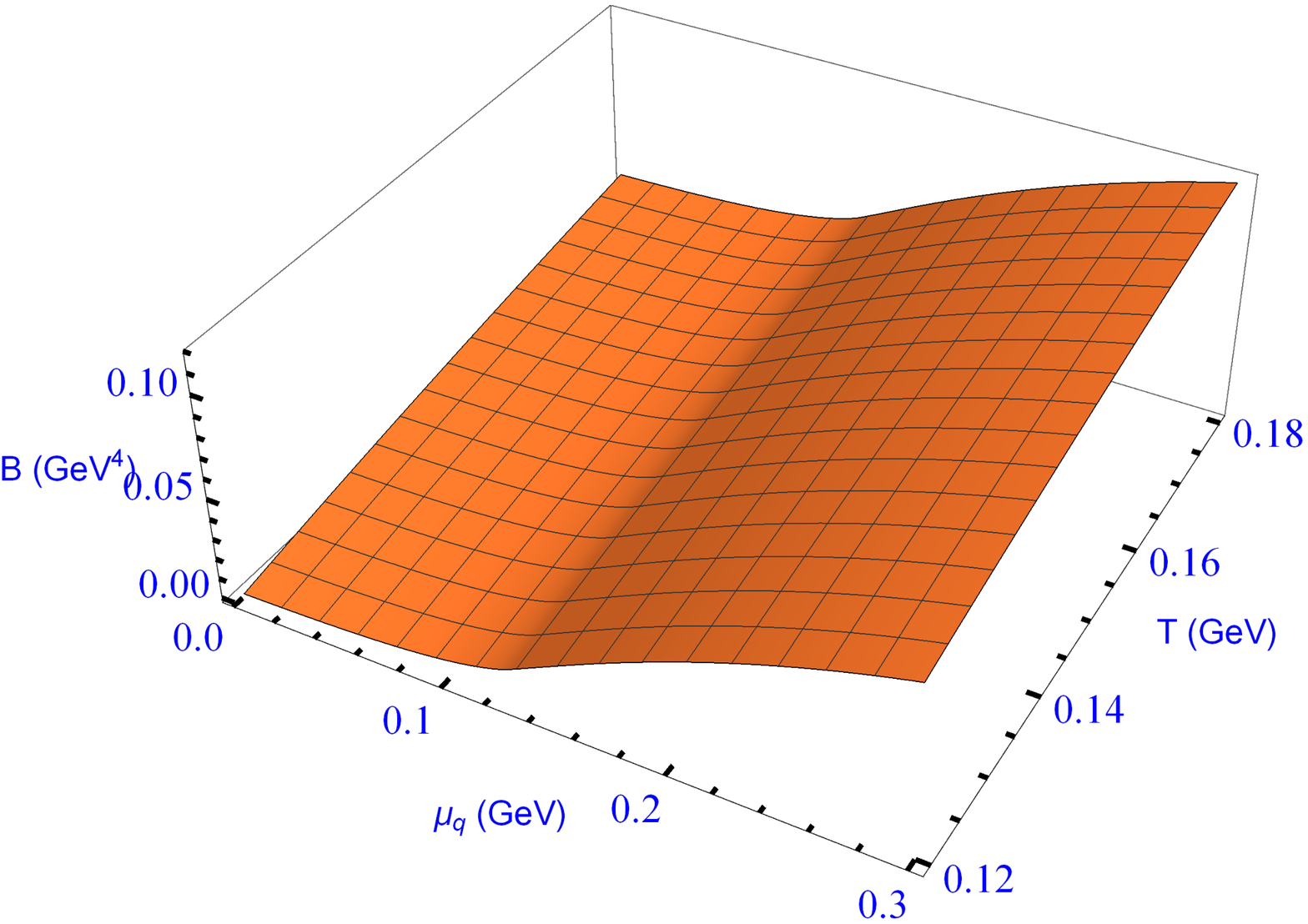}}
\caption{The derived quasiparticle mass in the parameter space according to the form given by Eq.~\eqref{ansaL2} and the fit shown in Fig.~\ref{massud2}.
(a) The quasiparticle mass $m$ as a function of $T$ and $\mu$ at $k = 1$ GeV.
(b) The quasiparticle mass $m$ as a function of $k$ and $\mu$ at $T = 0.12$ GeV.
(c) The bag constant $B$ as a function of $T$ and $\mu$.}
\label{massfinal2}
\end{figure}

The resultant mass function $m(k, T, \mu)$ and the bag constant $B(T, \mu)$ are shown in Fig.~\ref{massfinal2}.
The left and middle plots give the mass as a function of $T$ and $\mu$ at given $k=1$ and as a function of $k$ and $\mu$ at given $T=0.12$.
Again, the bag constant can be obtained by numerical integration of Eqs.~\eqref{go2} and~\eqref{go3}.
The dependence of the bag constant on the temperature $T$ or chemical potential $\mu$ is presented in the right plot of Fig.~\ref{massfinal2}.
The mass function and the bag constant are found to be moderate in $T$ and $\mu$, mainly in accordance with the existing results~\cite{sph-eos-06}.
Different from Fig.~\ref{massfinal}, as $k\to 0$, the mass of the quasiparticle does not modify significantly.
Again, the obtained bag constant $B$ is manifestly path independent.

\begin{figure}[htb]
  \centering
  \subfigure[]{
    \includegraphics[scale=0.3]{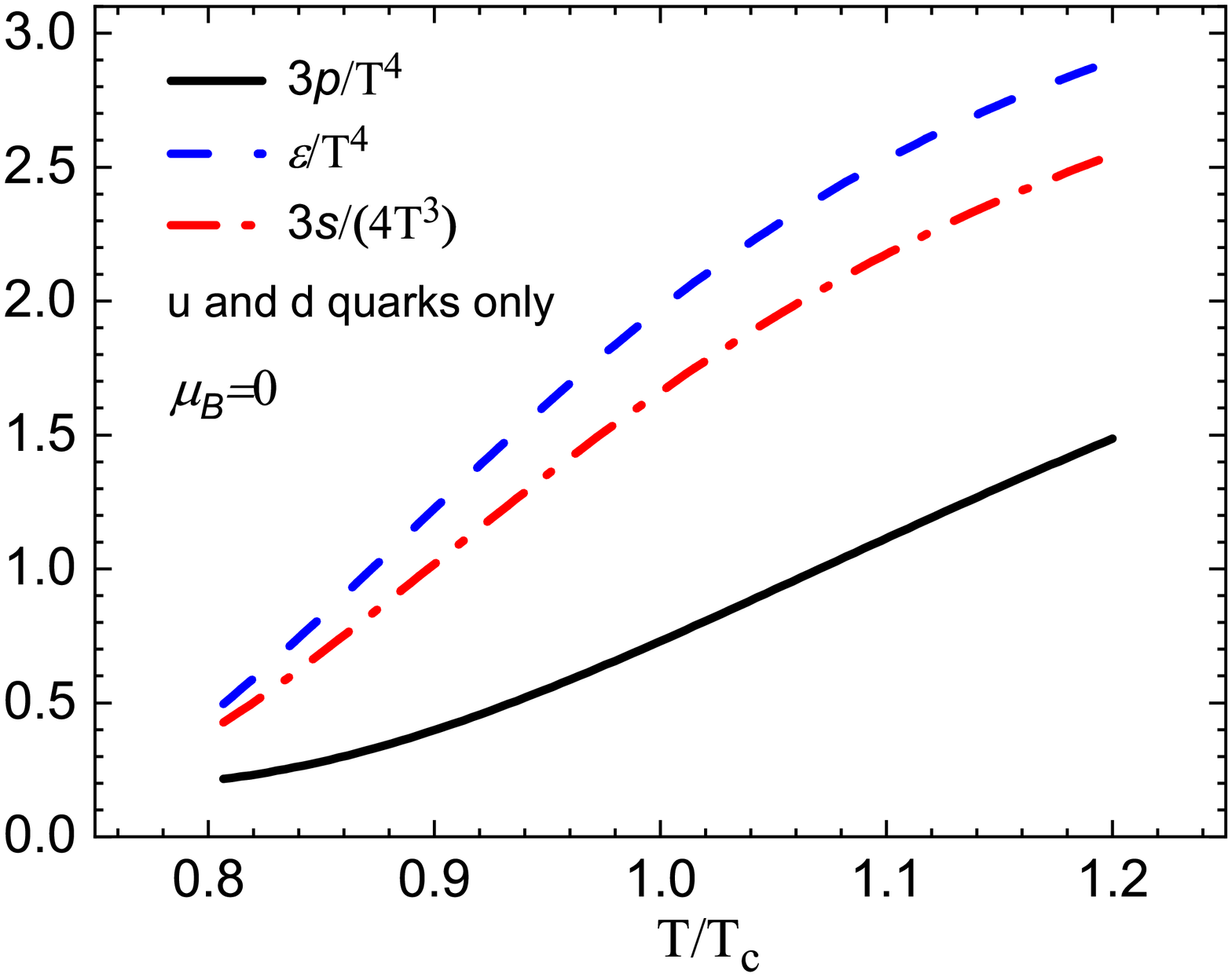}}
  \hspace{0.01in}
  \subfigure[]{
    \includegraphics[scale=0.3]{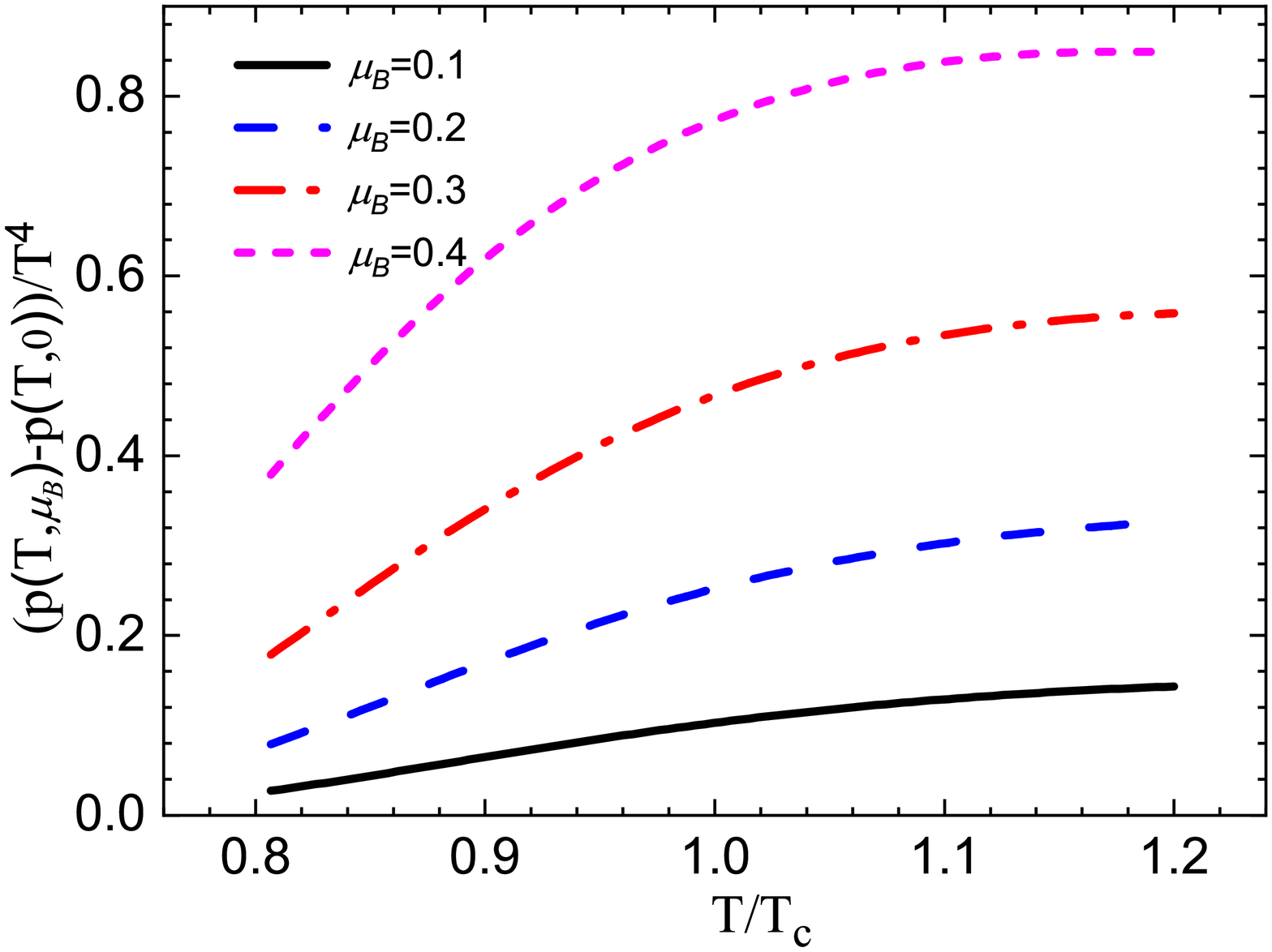}}
\caption{The derived thermodynamic quantities by considering the $u$ and $d$ quarks, where the $s$ quarks, gluons, and anti-particles are not explicitly taken into account.
(a) The pressure $3p/T^4$, energy density $\epsilon/T^4$, and entropy density $3s/(4T^3)$ as functions of the temperature $T/T_c$ at vanishing chemical potential, where $T_c=1.5$ GeV in accordance with lattice QCD data.
(b) The difference in pressure between the states with finite and vanishing chemical potential is shown as a function of temperature $T/T_c$. }
\label{therm}
\end{figure}

Before closing this section, we present in Fig.~\ref{therm} a few resulting thermodynamic quantities evaluated using the toy model proposed in case 2.
In the left plot, we show the pressure, energy density, and entropy density as a function of temperature at vanishing chemical potential.
The right plot gives the difference in pressure between the states with finite and vanishing chemical potential.
It is noted in the calculations, in accordance with the simplified scenario, one only takes into account the $u$ and $d$ quarks but does not include the $s$ quarks, gluons, or the anti-particles.
By comparing the results with those obtained using more sophisticated approaches~\cite{quasiparticle-latt-eos-34, quasiparticle-latt-eos-37, sph-eos-06}, one is led to the following observations.
The tendency of the temperature dependence is mainly correct, while the magnitudes of the calculated thermodynamic quantities consistently underestimate the existing results.
This is because the simplified models do not consider the contributions from the remaining degrees of freedom, including those from the anti-particles. 
Moreover, the order of magnitude for these quantities can be roughly recuperated by multiplying a factor of two.
The latter effectively compensates for the contributions missing from the anti-particles.
We note, nonetheless, that the main objective of the present approach is to explore the analytic properties of the mass function from a bottom-up perspective rather than reproduce the lattice data numerically by employing some sophisticated approximate function.

\section{Concluding remarks}\label{sec4}

To summarize, in this work, we reviewed the topic of the quasiparticle model closely related to Ru-Keng Su's distinguished contributions in the past years.
Moreover, we explore the approach applied to scenarios with finite chemical potential.
Different from the standard recipe in the literature, we explored the possibility that the effective mass of the quasiparticle might be a function of the momentum, in addition to the dependence on temperature and chemical potential.
It was shown that such a scenario emerges as a special solution to an integro-differential equation derived from the thermodynamic consistency.
We pointed out that the special solution in question is essentially a generalization to those previously explored in the literature.
Instead of fitting to the lattice QCD data at vanishing chemical potential, we performed a ``bottom-up'' approach by assuming two analytic ansatzes.
The remaining physical quantities were subsequently derived and discussed.
We note that the momentum-dependent quanta mass has also been addressed by some authors from the QCD perspective, where the analyses were closely related to the symmetry of the underlying system.
In terms of the Gribov-Zwanziger framework, results on the gluon~\cite{qcd-RGZ-01,qcd-RGZ-02,qcd-RGZ-04,qcd-RGZ-05} and quark propagator~\cite{qcd-GZ-02} indicated that the pole masses are functions of the momentum.
Besides, calculations using the Schwinger-Dyson equation showed momentum-dependence for both gluon~\cite{qcd-DSE-02} and quark~\cite{qcd-DSE-03,qcd-DSE-04} dynamic masses.

The current approach's main objective is to explore the analytic properties of the mass function.
It is primarily motivated as one might distinguish the various roots deriving from the thermodynamical consistency condition.
As observed and discussed in the main text, these different roots are somehow separated by the pole of the relevant equation, which is not apparent if a numerical scheme were utilized in the first place.
The calculations primarily employ Eq.~\eqref{go7b}.
It is a simplified approach as it ignores anti-particles' contributions and is only utilized to fit to accommodate the u and d quarks.
On the other hand, a numerical approach directly based on Eq.~\eqref{gozero} was carried out in a previous study~\cite{sph-eos-06}, where the cancelations warranted by Eqs.~\eqref{go2} and~\eqref{go3} take place for individual particles, as well as their anti-particles.
Nonetheless, the present study gives rise to the following speculations.
First, we have attempted to avoid the singularity of the mass function by entirely evading its poles by imposing the conditions, Eqs.~\eqref{ansaCon1} and~\eqref{ansaCon3}.
The resultant physical quantities are, in turn, manifestly {\it analytic} on the $T$ and $\mu$ parameter space.
Curiously, from a theoretical perspective, one expects a curve of first-order phase transition on the parameter plane, which entails some discontinuity.
In other words, the discontinuity avoided in the present study might be utilized in our favor.
Specifically, a pole in the mass function indicates an infinite mass, which can be viewed as a natural and benign outcome when a degree of freedom can hardly be excited.
We plan to address these aspects in further studies.

\section*{Acknowledgements}

This work is supported by the National Natural Science Foundation of China.
This work is partially supported by the Central Government Guidance Funds for Local Scientific and Technological Development, China (No. Guike ZY22096024).
We also gratefully acknowledge the financial support from Brazilian agencies 
Funda\c{c}\~ao de Amparo \`a Pesquisa do Estado de S\~ao Paulo (FAPESP), 
Funda\c{c}\~ao de Amparo \`a Pesquisa do Estado do Rio de Janeiro (FAPERJ), 
Conselho Nacional de Desenvolvimento Cient\'{\i}fico e Tecnol\'ogico (CNPq), 
and Coordena\c{c}\~ao de Aperfei\c{c}oamento de Pessoal de N\'ivel Superior (CAPES).

\bibliographystyle{h-physrev}
\bibliography{references_qian}

\end{document}